\documentclass[12pt,fleqn]{article}
\usepackage{epsfig,times,latexsym,enumerate,lscape}
\newcommand{\prd}{Phys. Rev. D }
\newcommand{\pp}{Preprint gr-qc/}

\newcommand{\prl}{Phys. Rev. Lett. }
\begin{document}

\title{Data analysis of continuous gravitational wave: Fourier transform-II}
\author{D.C. Srivastava$^{1,2}$\thanks{e-mail: dcsrivastava@now-india.com} 
      $\,$ and S.K. Sahay$^1$\thanks{e-mail: ssahay@iucaa.ernet.in}\thanks{Present address: Inter University Centre for Astronomy and Astrophysics, Post Bag 4, Ganeshkhind, Pune - 411007, India}\\ \\
\normalsize $^1$Department of Physics, DDU Gorakhpur University, Gorakhpur-273009, U.P., India.\\
\normalsize $^2$Visiting Associate, Inter University Centre for Astronomy and Astrophysics, Post  \\
\normalsize Bag 4, Ganeshkhind, Pune-411007, India.}
\date{}
\maketitle

\begin{abstract}
In this paper we  obtain the  Fourier Transform of a continuous gravitational
wave. We have analysed the data set for (i) one year observation time and (ii)
arbitrary observation time, for arbitrary location of detector and source
taking into account the effects arising due to rotational as well as orbital
motion of the earth. As an application of the transform we considered spin down and
N-component signal analysis.
\end{abstract}
\section{Introduction}
Detection of gravitational waves (GW) mainly depends on the efficient Fourier analysis
of the data of the output of the detector such as LIGO/VIRGO. In an earlier 
paper we presented Fourier analysis of one day observation data set of the
response of a laser interferometer (Srivastava and Sahay, 2002 a). Hereafter
the reference to this paper and its results are noted by the prefix I. 
We have seen that the
amplitude and the frequency modulations result into a large number of side bands
about the signal frequency $f_o$. Consequently, the maximum power lies in the
frequency $f_o + 2f_{rot}$ with amplitude reduction by 74\% to what one would
have expected due to increased data interval. Here $f_{rot}$ is the rotational
frequency of the earth. Hence, for GW detection it is
desirable to obtain FT for longer observation data. In the next
section we investigate one year observation data. Incidently there exists correspondences
and identifications of the present analysis to the analysis of paper-I. To facilitate 
analogous modifications we introduce corresponding quantities with tilde viz., 
$\tilde{\cal C}$ and $\tilde{\cal D}$ in place of ${\cal C}$ and ${\cal D}$.
We obtain in section 3 generalisations of the results for arbitrary observation time. 
As an application of the results obtained we consider spin
down and N-component signal analysis in sections 4 and 5. Last section contains
the discussion and conclusions of the paper.

\section{Fourier transform for one year integration}
\label{sec:fty}
The phase of the GW signal of frequency $f_o$ at time $t$ is given via (I-25) i.e. 
\begin{eqnarray}
\Phi(t) & = & 2\pi f_o t + {\cal Z}\cos (w_{orb}t - \phi ) + 
{\cal N}\cos (w_{rot}t - \delta ) - {\cal R} - {\cal Q}
\end{eqnarray}

\noindent where \\

\begin{equation}
\label{eq:daypq}
\left.\begin{array}{lcl}
\vspace{0.2cm}
{\cal P}& = & 2\pi f_o {R_e\over c} \sin\alpha (\cos\beta _o(\sin
\theta \cos\epsilon \sin\phi + \cos\theta \sin\epsilon )
 - \sin\beta _o \sin\theta \cos\phi )\, ,\\
\vspace{0.2cm}
{\cal Q}& = & 2\pi f_o {R_e\over c}\sin\alpha (\sin\beta _o (\sin\theta \cos
\epsilon \sin\phi + \cos\theta \sin\epsilon )
+ \cos\beta _o \sin\theta\cos\phi ) \, ,\\
\vspace{0.2cm}
{\cal N}& = & \sqrt{ {\cal P}^2 + {\cal Q}^2 }\, ,\\
\vspace{0.2cm}
{\cal Z}& = & 2\pi f_o {R_{se}\over c}\sin\theta \, , \\
{\cal R}& = & {\cal Z}\cos\phi\, ,\\
\end{array} \right\}
\end{equation}

\vspace{0.2in}

\begin{equation}
\label{eq:daydelta}
\left.\begin{array}{lcl}
\vspace{0.2cm}
\delta & = &  \tan^{- 1}\frac{{\cal P}}{{\cal Q}}\, ,\\
\vspace{0.2cm}
\phi' & = & w_{orb}t - \phi\, ,\\
\vspace{0.2cm}
\xi_{orb} & = & w_{orb}t\; = \; a\xi_{rot};\quad  a \;= \; w_{orb}/w_{rot}\; \approx \; 1/365.26\, , \\
\xi_{rot} & = & w_{rot}t 
\end{array} \right\}
\end{equation}

\noindent  where $R_{e}$, $R_{se}$, $w_{rot}$ and $w_{orb}$ represent respectively the
earth's radius, average distance of earth's centre from the origin of
Solar System Barycentre (SSB) frame, the rotational and the orbital angular velocity of the 
earth. $\theta$ and $\phi$ denote the celestial colatitude and celestial longitude
of the source. $\epsilon$ and $c$ represent the obliquity of the ecliptic and the velocity of
light. $\alpha$ is the colatitude of the detector. Here $t$ represents the
time in seconds elapsed from the instant the sun is at the Vernal Equinox and $\beta_o$
is the local sidereal time at that instant, expressed in radians. \\ 

\noindent The Fourier transform (FT) for one year observation time, $T_{obs}$ is given as

\begin{equation}
\left[\tilde{h}(f)\right]_y = \int_0^{\bar{a}T} \cos[\Phi (t)]e^{-i2\pi ft}dt\; ;
\label{eq:hfy1}
\end{equation}
\begin{equation}
\bar{a} = a^{-1} = w_{rot}/w_{orb}\; ;\quad T =\; one\; sidereal\; day \; ;
\end{equation}
\begin{equation}
T_{obs} = \bar{a}T \simeq  3.14 \times 10^{7}\; sec.
\end{equation}

\noindent This splits, similar to Eqs. (I-33-35), into two terms as
\begin{equation}
\left[\tilde{h}(f)\right]_y = I_{\bar{\nu}_-} + I_{\bar{\nu}_+} \; ;
\label{eq:inuy}
\end{equation}
\begin{eqnarray}
\label{eq:inubar}
 I_{\bar{\nu}_-}& =& {1\over 2 w_{orb}}\int_0^{2\pi} e^{i \left[
 \bar{\xi}\bar{\nu}_- +
{\cal Z}\cos (\bar{\xi} - \phi ) + {\cal N}\cos (\bar{a}\bar{\xi} - \delta ) -{\cal R}  - {\cal 
Q} \right] } d\bar{\xi}\, ,\\
I_{\bar{\nu}_+}& = & {1\over 2 w_{orb}}\int_0^{2\pi} e^{- i \left[
\bar{\xi}\bar{\nu}_+
+ {\cal Z} \cos (\bar{\xi} - \phi ) + {\cal N}\cos (\bar{a}\bar{\xi} - \delta ) - {\cal R} - {
\cal Q} \right] } d\bar{\xi}\, , \\
\bar{\nu}_{\mp}& = & \frac{f_o \mp f}{f_{orb}} ; \quad \bar{\xi} \; = \; \xi_{orb} \; = \;
w_{orb}t  
\end{eqnarray}

\noindent where $f_{orb}$ is the orbital frequency of the earth. Hereafter, we neglect 
$I_{\bar{\nu}_+}$ as it oscillates rapidly and contributes very
little to $\left[\tilde{h}(f)\right]_y$ and write $\bar{\nu}$ in place of $\bar{\nu}_-$. A 
careful comparison of
Eq.~(\ref{eq:inubar}) with (I-33) i.e. 
\begin{equation}
\label{eq:dayinu}
 I_{\nu_-} = {1\over 2 w_{rot}}\int_0^{2\pi} e^{i \left[ \xi\nu_- +
{\cal Z}\cos (a\xi - \phi ) + {\cal N}\cos (\xi - \delta ) -{\cal R}  - {\cal 
Q} \right] } d\xi
\end{equation}

\noindent obtained for one day observation data reveals that the integrand of the
equations are identical with following identifications and correspondences.

\begin{equation}
\left.\begin{array}{ccc}
\delta &\leftrightarrow &\phi\; ,\\
{\cal Z}&\leftrightarrow &{\cal N}\; , \\
a &\leftrightarrow & \bar{a}.
\end{array} \right\}
\end{equation}

\noindent We may, therefore, employ the results obtained there by introducing obvious
corresponding quantities i.e. $\bar{\cal B}$, $\bar{\cal C}$, $\bar{\cal D}$
in place of ${\cal B}$, ${\cal C}$, ${\cal D}$ leaving ${\cal A}$ unchanged. Hence

\begin{equation}
\label{eq:hfy}
\left[\tilde {h}(f)\right]_y  \simeq  \frac{\bar{\nu}}{2 w_{orb}} \sum_{k  =  - \infty}^{k =  
\infty} \sum_{m = - \infty}^{m =  \infty} e^{ i {\cal A}}\bar{{\cal B}}[
\bar{{\cal C}} - i\bar{{\cal D}} ]\; ;
\end{equation}  

\begin{equation}
\left.\begin{array}{lcl}
\vspace{0.2cm}
{\cal A}&  = &{(k + m)\pi\over 2} - {\cal R} - {\cal Q} \; ,\\
\vspace{0.2cm}
\bar{{\cal B}} & = & {J_k({\cal N}) J_m({\cal Z})\over {\bar{\nu}^2 - (\bar{a} k + m)^2}} \; , \\
\vspace{0.2cm}
\bar{{\cal C}} &= & \sin 2\bar{\nu}\pi \cos ( 2 \bar{a} k \pi - k \delta - m \phi ) -  
{ \bar{a} k + m \over \bar{\nu}}\{\cos 2 \bar{\nu} \pi \sin ( 2 \bar{a} k \pi - k \delta - m \phi )
+ \sin ( k \delta + m \phi )\}\; , \\
\vspace{0.2cm}
\bar{{\cal D}} & = & \cos 2\bar{\nu}\pi \cos ( 2 \bar{a} k \pi - k \delta - m \phi ) + 
 {k \bar{a} +m \over \bar{\nu}}\sin 2 \bar{\nu} \pi \sin ( 2 \bar{a} k \pi - k \delta - m \phi )
 - \cos ( k \delta + m \phi )  
\end{array}\right\}
\end{equation}

\noindent where $J$ stands for the Bessel function of first kind. Now the FT of the two 
polarisation states can be written as

\begin{eqnarray}
\label{eq:hfpy}
\left[\tilde{h}_+(f)\right]_y&=&h_{o_+}\left[\tilde{h}(f)\right]_y \nonumber \\
&\simeq & \frac{\bar{\nu} h_{o_+}}{2 w_{orb}} \sum_{k  =  - \infty}^{k =  
\infty} \sum_{m = - \infty}^{m =  \infty} e^{ i {\cal A}}\bar{{\cal B}}[
\bar{{\cal C}} - i\bar{{\cal D}} ]\,\;
\end{eqnarray}
and
\begin{eqnarray}
\label{eq:hfcy}
\left[\tilde{h}_\times(f)\right]_y&=& - i h_{o_\times}\left[\tilde{h}(f)\right]_y \nonumber \\
&\simeq & \frac{\bar{\nu} h_{o_\times}}{2 w_{orb}} \sum_{k  =  - \infty}^{k =  
\infty} \sum_{m = - \infty}^{m =  \infty} e^{ i {\cal A}}\bar{{\cal B}}[
\bar{{\cal D}} - i\bar{{\cal C}} ]
\end{eqnarray}

\par The FT obtained contains double series of the Bessel
functions of the orders $k$ and $m$ ranging from $-\infty$ to $\infty$. It is 
well known that the Bessel functions decrease rapidly as the order exceeds the
argument. Hence possible range of $k$ and $m$ over which the summation of the series is
to be considered depends on the arguments of Bessel functions i.e ${\cal Z}$ and 
${\cal N}$. Referring to Eq.~(\ref{eq:daypq}) it is found
that

\begin{equation}
\left.\begin{array}{ccl}
\vspace{0.2cm}
{\cal Z}_{max} & = &  3133215\left(\frac{f}{1\, KHz} \right)\\
{\cal N}_{max} & = &134\left(\frac{f}{1\, KHz} \right)
\end{array}\right\}
\end{equation}

\begin{figure}
\centering
\epsfig{file=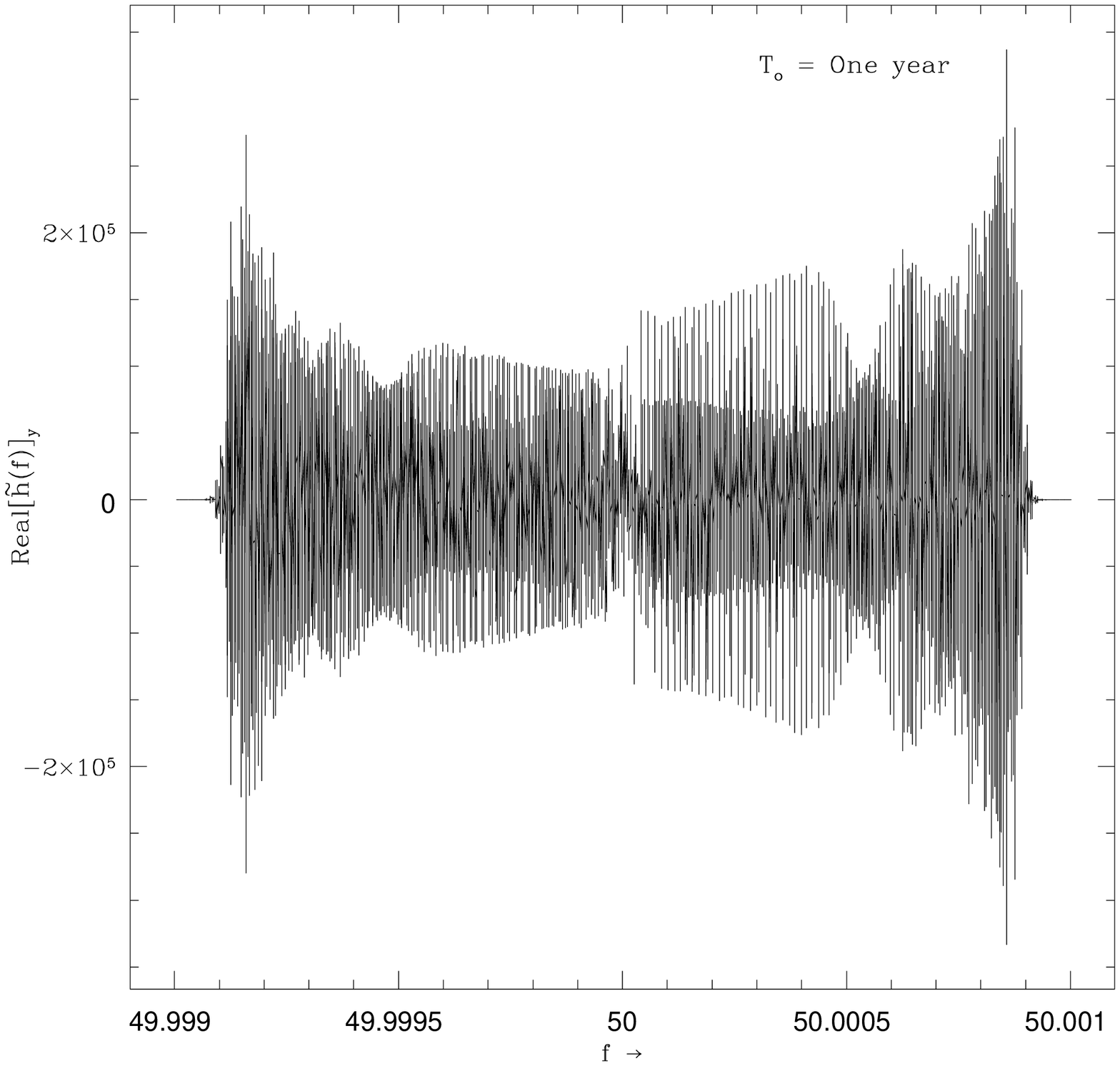,height=16.0cm}
\caption{FT of a FM signal of frequency, $f_o = 50$ Hz from a source located at $(\pi /18 , 0)$
with a resolution of $3.17 \times 10^{-8}.$ }
\label{fig:yrealfm}
\end{figure}

\noindent The FT of a FM signal for

\begin{equation}
\left.\begin{array}{lll}
\vspace{0.2cm}
f_o = 50\; Hz\, , & h_{o_+} = h_{o_\times} = 1 & \\
\vspace{0.2cm}
\alpha = \pi /4\, , & \beta_o = 0\, , & \gamma =  \pi \, , \\
\theta = \pi /18\, , & \phi = 0\, , &  \psi = \pi /4.
\end{array} \right\}
\label{eq:yearloc}
\end{equation}

\noindent is shown in Fig.~(\ref{fig:yrealfm}). The spectrum 
resolution is kept equal to $1/T_o$ i.e. $ 3.17 \times 10^{-8}$ Hz. We have
convinced ourselves by plotting the FT at higher resolutions that the
resolution of $1/T_o$ is sufficient to represent the relevant
peaks. We notice that the drop in amplitude at the source frequency is about
98\% which may be
attributed to the presence of a large number of side bands.\\

\par The complete response $R(t)$ of the detector and its FT
can be obtained similar to the one presented in section 4 of the paper-I.
Taking the results straight away from (I-43, 44, 45) one may get

\begin{eqnarray}
\label{eq:rfy}
\left[\tilde {R}(f)\right]_y &= & \left[\tilde{R}_+(f)\right]_y +
\left[\tilde{R}_\times (f)\right]_y \nonumber \\
 & = & e^{-i 2 \beta_o}\left[\tilde{h}( f
+ 2 f_{rot})/2\right]_y\left[ h_{o_+}( F_{1_+} + i F_{2_+} )
+ h_{o_\times} ( F_{2_\times} - i F_{1_\times} )\right] +\nonumber \\
&& e^{i2\beta_o}\left[\tilde{h}( f - 2 f_{rot})/2\right]_y\left[ h_{o_+}
( F_{1_+} - i F_{2_+} )
- h_{o_\times} ( F_{2_\times} + i F_{1_\times} )\right] +\nonumber \\
& & e^{-i\beta_o}\left[\tilde{h}( f + f_{rot})/2\right]_y\left[ h_{o_+}( F_{3_+} + i F_{4_+} )
+ h_{o_\times} ( F_{4_\times} - i F_{3_\times} )\right] +\nonumber \\
&& e^{i\beta_o}\left[\tilde{h}( f - f_{rot})/2\right]_y\left[ h_{o_+}( F_{3_+} - i F_{4_+} )
- h_{o_\times} ( F_{4_\times} + i F_{3_\times} )\right] +\nonumber \\
&&\left[\tilde{h}(f)\right]_y\left[ h_{o_+}F_{5_+} - i h_{o_\times}F_{5_\times}\right]
\end{eqnarray}

\par
\begin{figure}
\centering
\epsfig{file=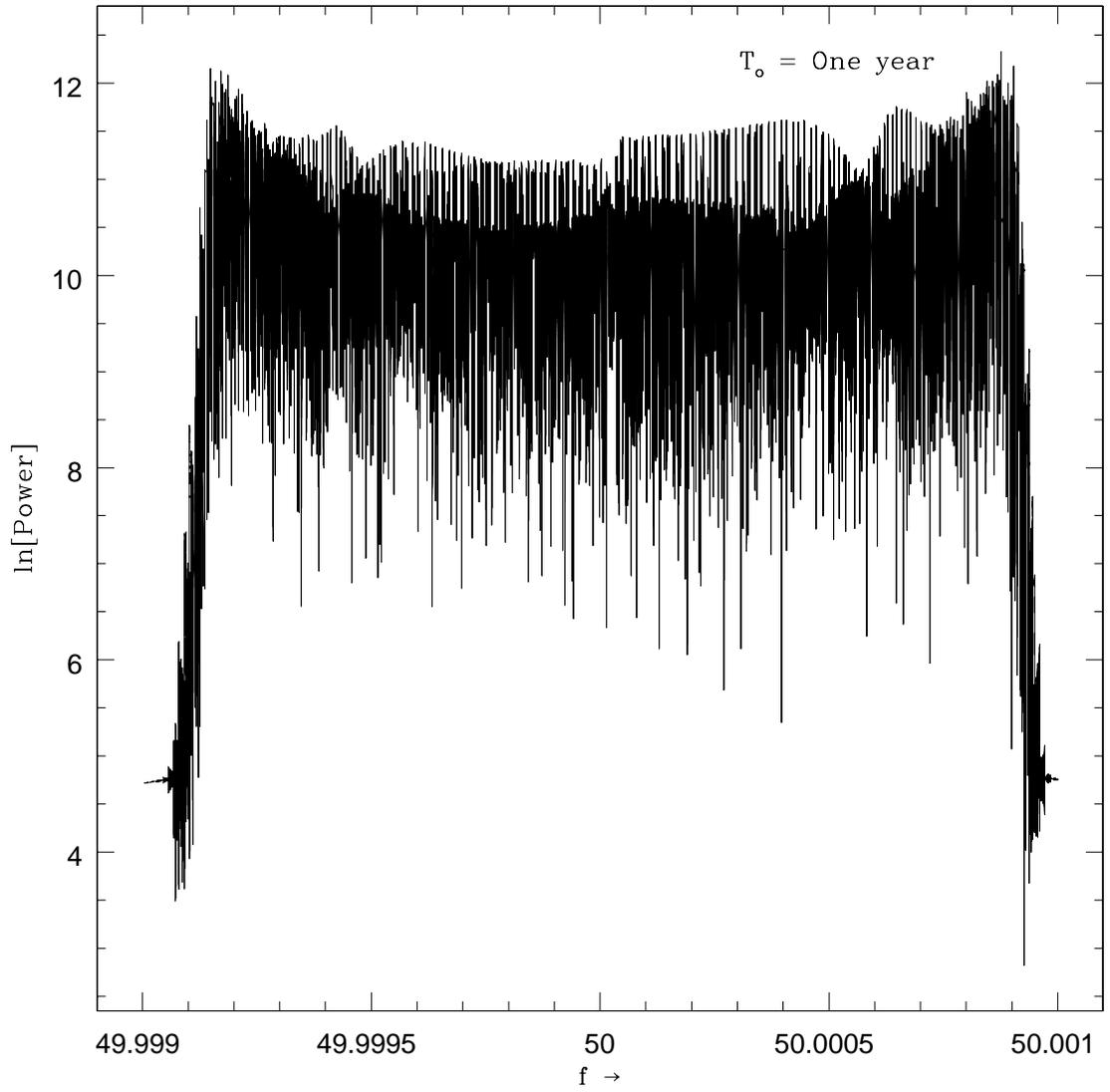,height=16.0cm}
\caption{Power spectrum of the complete response of a modulated signal of frequency, $f_o = 50$ 
Hz from a source located at $ (\pi /18 ,0)$ with a resolution of $3.17 \times 10^{-8}.$}
\label{fig:ycr}
\end{figure}

 Figure~(\ref{fig:ycr}) shows the power spectrum of the complete response of
the Doppler modulated signal. We have kept here all the parameters same as in FM. 
In this caee also we observe that maximum  
power is associated with $f_o + 2f_{rot}$ and the least lies in $f_o - f_{rot}$.
\section{Fourier transform for arbitrary observation time}
\label{sec:ftt}
It is important to obtain the FT for arbitrary observation 
time. The results obtained will be
employed to outline how spin down of a pulsar due to the gravitational
radiation back reaction or due to some other mechanism can be taken into account.

\noindent The FT for a data of observation time $T_o$ is given via

\begin{equation}
\tilde{h}(f) \;= \;\int_0^{T_o}\cos [\Phi (t)]e^{-i2\pi f t} dt
\end{equation}

\noindent which may be split

\begin{equation} \tilde{h}(f)\ = I_{\nu_-} + I_{\nu_+} \; ;
 \label{eq:inut1}
 \end{equation}
 \begin{eqnarray}
\label{eq:inut}
 I_{\nu_-}& =& {1\over 2 w_{rot}}\int_0^{\xi_o} e^{i \left[
 \xi\nu_- +
{\cal Z}\cos (a\xi - \phi ) + {\cal N}\cos (\xi - \delta ) -{\cal R}  - {\cal 
Q} \right] } d\xi\, ,\\
I_{\nu_+}& = & {1\over 2 w_{rot}}\int_0^{\xi_o} e^{- i \left[
\xi\nu_+
+ {\cal Z} \cos (a\xi - \phi ) + {\cal N}\cos (\xi - \delta ) - {\cal R} - {
\cal Q} \right] } d\xi\, , \\
\nu_{\mp}& = & \frac{f_o \mp f}{f_{rot}} ; \quad \xi_o \; = \; w_{rot}T_o ;
\quad \xi \; = \; \xi_{rot} \; = \; w_{rot}t \, , 
\end{eqnarray}

\noindent [refer to Eqs. I (33 - 35)].

\noindent As $I_{\nu_+}$ contributes very little to $\tilde{h}(f)$
we drop $I_{\nu_+}$ and write $\nu$ in place of $\nu_-$. Using 
the identity

\begin{equation}
e^{\pm i\kappa\cos\vartheta} = J_o (\pm\kappa) + 2 \sum_{l = 1}^{l = 
\infty} i^l J_l (\pm\kappa)\cos l\vartheta
\label{eq:bessel}
\end{equation}

\noindent we obtain 
\begin{eqnarray}
\tilde{h}(f) &\simeq & \frac{1}{2 w_{rot}} e^{i ( 
- {\cal R} - {\cal Q})} \int_0^{\xi_o} e^{i\nu\xi} \left[ J_o( {\cal Z} ) + 2 
\sum_{k = 1}^{k =  \infty} J_k ({\cal Z}) i^k \cos k (a\xi - \phi )\right] 
 \nonumber \\
&& \times\,\left[J_o( {\cal N} ) + 2 \sum_{m = 1}^{m =  \infty} J_m ({\cal N}) i^m 
\cos m (\xi - \delta )\right] d\xi
\end{eqnarray}

\noindent After performing the integration and proceeding in a straight-forward
manner we have

\begin{equation}
\label{eq:hft}
\tilde{h}(f) \simeq  \frac{\nu}{2 w_{rot}} \sum_{k  =  - 
\infty}^{k = \infty} \sum_{m = - \infty}^{m =  \infty} e^{ i {\cal A}}{\cal 
B}[ \tilde{{\cal C}} - i\tilde{{\cal D}} ] \; ; \;
\end{equation}  

\begin{equation}
\left.\begin{array}{lcl}
\vspace{0.2cm}
{\cal A}&  = &{(k + m)\pi\over 2} - {\cal R} - {\cal Q}  \\
\vspace{0.2cm}
{\cal B} & = & {J_k({\cal Z}) J_m({\cal N})\over {\nu^2 - (a k + m)^2}} \\
\vspace{0.2cm}
\tilde{{\cal C}} &=& \sin \nu\xi_o \cos ( a k \xi_o + m\xi_o - k \phi - m \delta ) -  
{ a k + m \over \nu}\{\cos\nu\xi_o \sin ( a k \xi_o + m\xi_o - k \phi - m \delta ) \nonumber \\
\vspace{0.2cm}
&&+ \sin ( k \phi + m \delta )\}\nonumber \\
\vspace{0.2cm}
\tilde{{\cal D}} & = & \cos \nu\xi_o \cos ( a k \xi_o + m\xi_o - k \phi - m \delta ) + 
{k a + m \over \nu}\sin \nu \xi_o \sin ( a k \xi_o + m\xi_o - k \phi - m \delta ) \nonumber \\
\vspace{0.2cm}
&& - \cos ( k \phi + m \delta )
\end{array} \right\}
\end{equation}

\noindent The FT of the two polarisation states of the wave can now be written as
\begin{eqnarray}
\label{eq:hfpt}
{h}_+(f)&=&h_{o_+}\tilde{h}(f) \nonumber \\
&\simeq & \frac{\nu h_{o_+}}{2 w_{rot}} \sum_{k  =  - \infty}^{k =
\infty} \sum_{m = - \infty}^{m =  \infty} e^{ i {\cal A}}{\cal B}[ \tilde{{\cal C}}
- i\tilde{{\cal D}} ] \; ;
\end{eqnarray}
\begin{eqnarray}
\label{eq:hfct}
\tilde{h}_\times (f) &=&- i h_{o_\times}\tilde{h}(f) \nonumber \\
&\simeq & \frac{\nu h_{o_\times}}{2 w_{rot}} \sum_{k  =  - \infty}^{k =  
\infty} \sum_{m = - \infty}^{m =  \infty} e^{ i {\cal A}}{\cal B}[ \tilde{{\cal D}}
- i\tilde{{\cal C}} ]
\end{eqnarray}

\noindent Now it is simple matter to obtain the FT of complete response. One gets
\begin{eqnarray}
\label{eq:rft}
\tilde{R}(f) & = & e^{-i 2 \beta_o}\tilde{h}( f
+ 2 f_{rot})/2\left[ h_{o_+}( F_{1_+} + i F_{2_+} )
+ h_{o_\times} ( F_{2_\times} - i F_{1_\times} )\right] +\nonumber \\
&& e^{i2\beta_o}\tilde{h}( f - 2 f_{rot})/2\left[ h_{o_+}
( F_{1_+} - i F_{2_+} )
- h_{o_\times} ( F_{2_\times} + i F_{1_\times} )\right] +\nonumber \\
& & e^{-i\beta_o}\tilde{h}( f + f_{rot})/2\left[ h_{o_+}( F_{3_+} + i F_{4_+} )
+ h_{o_\times} ( F_{4_\times} - i F_{3_\times} )\right] +\nonumber \\
&& e^{i\beta_o}\tilde{h}( f - f_{rot})/2\left[ h_{o_+}( F_{3_+} - i F_{4_+} )
- h_{o_\times} ( F_{4_\times} + i F_{3_\times} )\right] +\nonumber \\
&&\tilde{h}(f)\left[ h_{o_+}F_{5_+} - i h_{o_\times}F_{5_\times}\right]
\end{eqnarray}

\begin{figure}
\centering
\epsfig{file=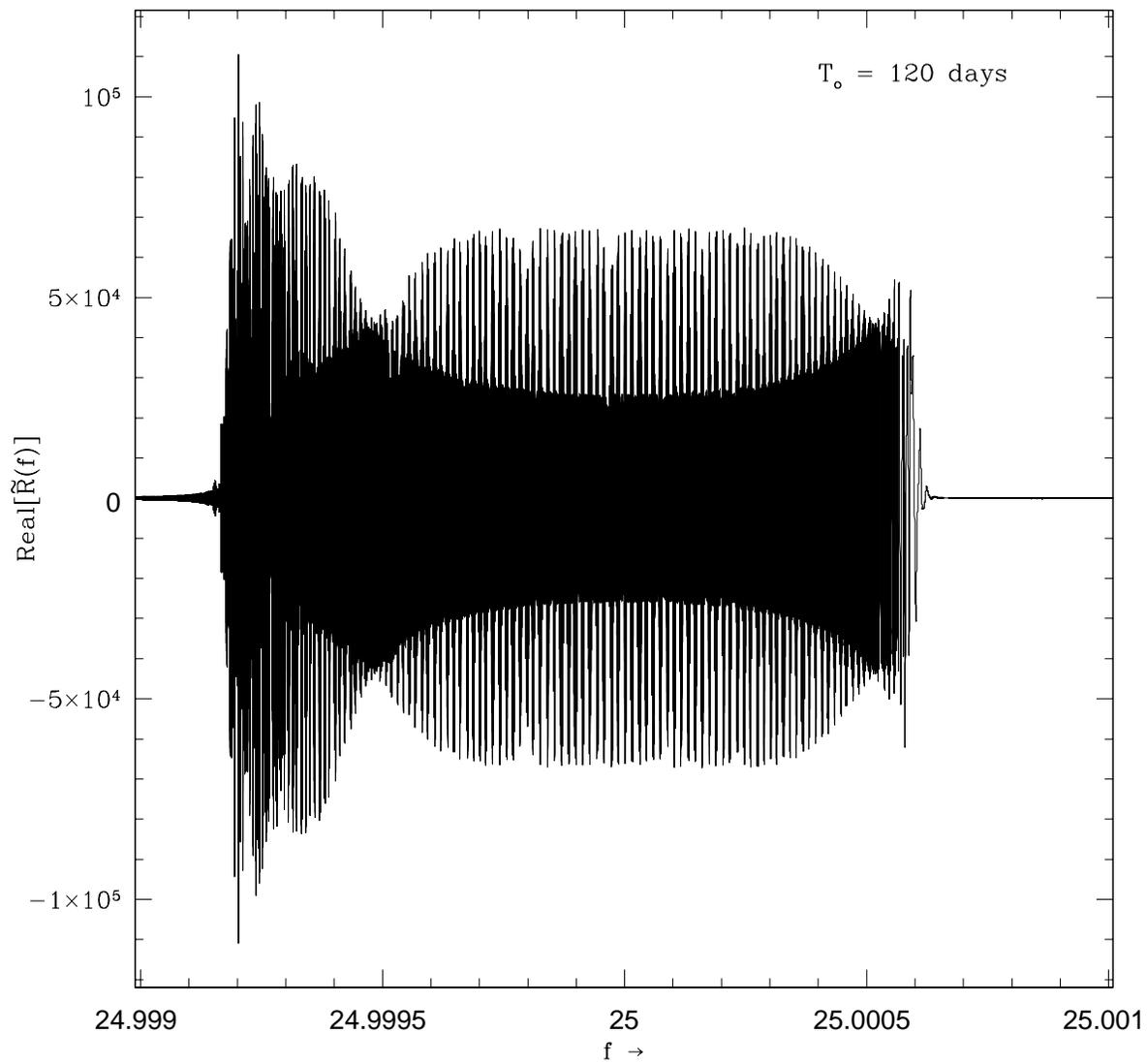,height=16.0cm}
\caption{FT of a FM signal of frequency, $f_o = 25$ Hz from a source located at $(\pi /9 , \pi /4)$
with a resolution of $9.67 \times 10^{-8}.$}
\label{fig:120realfm}
\end{figure}

\begin{figure}
\centering
\epsfig{file=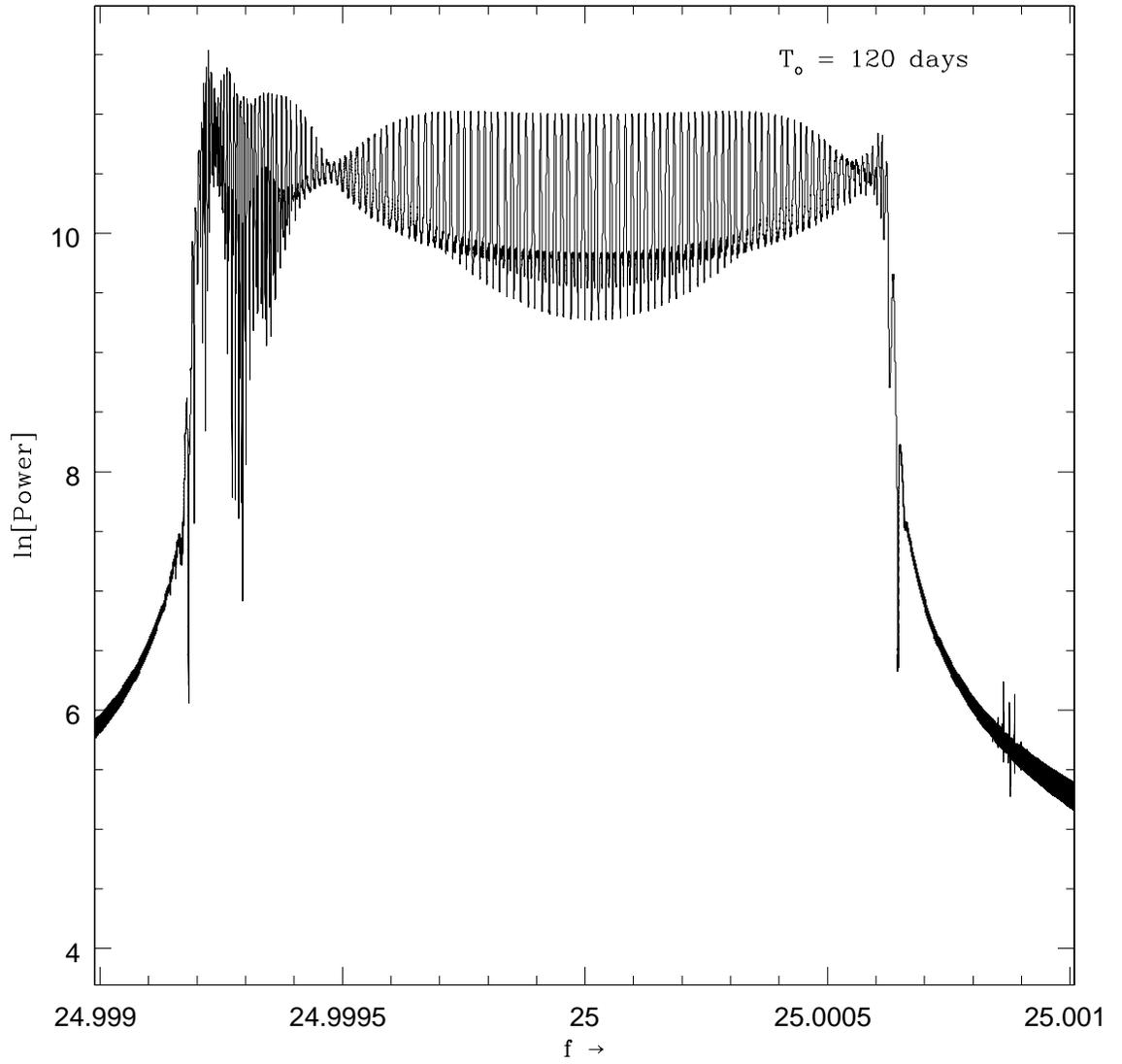,height=16.0cm}
\caption{Power spectrum of the complete response of a Doppler modulated signal of frequency, 
$f_o = 25$ Hz from a source located at $(\pi /9 , \pi /4)$ with a resolution of 
$9.67 \times 10^{-8}.$}
\label{fig:120cr}
\end{figure}

\noindent The FT of FM signal of a detector for a data of 120 days for

\begin{equation}
\left.\begin{array}{lll}
\vspace{0.2cm}
f_o = 25\; Hz\, , & h_{o_+} = h_{o_\times} = 1 & \\
\vspace{0.2cm}
\alpha = \pi /6\, , & \beta_o = \pi /3\, , & \gamma =  2\pi /3 \, , \\
\theta = \pi /9\, , & \phi = \pi /4\, , &  \psi = \pi /4. \\
\end{array} \right\}
\label{eq:120loc}
\vspace{0.4cm}
\end{equation}

\noindent is plotted in Fig.~(\ref{fig:120realfm}) 
with a resolution of $\,1/2T_o \approx 9.67 \times 10^{-8}$ Hz. The Power
spectrum of the complete response is plotted in Fig.~(\ref{fig:120cr}). In
this case also the most of the power of the signal lies in the frequency $f_o + 2f_{rot}$ 
and the least with $f_o - f_{rot}.$
\section{Spin down}
\label{sec:fts}
Pulsars loose their rotational energy by processes like electro-magnetic breaking,
emission of particles and emission of GW. Thus, the rotational 
frequency is not completely stable, but varies over a time scale which is of the 
order of the age of the pulsar. Typically, younger pulsars have the largest spin
down rates. Current observations suggest that spin down is primarily due to
electro-magnetic breaking (Manchester, 1992 and Kulkarni, 1992).
Over the entire observing time, $T_o$ the frequency drift would be small
but it may be taken into account for better sensitivity.   
To account this aspect we consider the evaluation of FT in a sequence of 
time windows by splitting the interval $0-T_o$ in M equal
 parts, each of interval $\bigtriangleup t$ $(T_o = M\bigtriangleup t$) 
such that the signal over a window may be treated as monochromatic. The
strategy is to evaluate the FT over the window and finally to add the result.
This process has been suggested by Brady and Creighton (2000) and Schutz (1998)
in numerical computing and
has been termed as {\it stacking\/} and {\it tracking\/}. For any such window
the initial and final times may be taken respectively as $t_o + n\bigtriangleup t$ and 
$t_o + (n + 1)\bigtriangleup t$ where $t_o$ represents the starting of the data
set and $0 \le n \le M - 1$.
The window under consideration is the $n_{th}$ window. Let
\begin{eqnarray}
I& =& \int_{t_o + n\bigtriangleup t}^{t_o + (n+1)\bigtriangleup t} h(\bar{t})
e^{-i2\pi f\bar{t}}d\bar{t} \nonumber \\
\label{eq:s1}
& = &\int_0^{\bigtriangleup t} h(t+t_o\;+
n\bigtriangleup t ) e^{-i2\pi (t+t_o+n\bigtriangleup t )f}dt\; ;\\
\bar{t} & = & t + t_o + n\bigtriangleup t
\label{eq:s2}
\end{eqnarray}

\noindent Hence the FT for the data of the time-window under consideration is given via

\begin{equation}
\label{eq:s3}
\left[\tilde{h}(f)\right]_s \;= \;\int_0^{\bigtriangleup t}\cos [\Phi (t + t_o
+ n\bigtriangleup t )]e^{-i2\pi (t + t_o + n\bigtriangleup t ) f} dt
\end{equation}

\noindent Taking the initial time of the data set
\begin{equation}
\label{eq:s4}
t_o = 0
\end{equation}

\noindent and proceeding as in the previous section we obtain 
\begin{eqnarray}
\left[\tilde{h}(f)\right]_s &\simeq & \frac{1 }{2 w_{rot}} e^{i \left[
2\pi n(f_o - f)\bigtriangleup t 
- {\cal R} - {\cal Q}\right]} \int_0^{\bigtriangleup t}
e^{i\nu\xi} \left[ J_o( {\cal Z} ) + 
2\sum_{k = 1}^{k =  \infty} J_k ({\cal Z}) i^k \cos k (a\xi - \lambda )\right] 
\;\times \nonumber \\
&& \left[J_o( {\cal N} ) + 2 \sum_{m = 1}^{m =  \infty} J_m ({\cal N}) i^m
\cos m (\xi - \zeta )\right] d\xi
\end{eqnarray}

\noindent After integration we get
\begin{eqnarray}
\label{eq:hfs}
\left[\tilde {h}(f)\right]_s& =& {\nu \over 2 w_{rot}} \sum_{k  =  - \infty}^{k =  
\infty} \sum_{m = - \infty}^{m =  \infty} e^{ i {\cal A}_s}{\cal B} \left[
{\cal C }_s  -  i {\cal D}_s \right]\, ; 
\end{eqnarray}
\begin{equation}
\left.\begin{array}{lcl}
\vspace{0.2cm}
{\cal A}_s&  = &{(k + m)\pi\over 2} + 2\pi n\bigtriangleup t(f_o - f)  - {\cal R} - {\cal Q}\\
\vspace{0.2cm}
{\cal B} & = & {J_k({\cal Z}) J_m({\cal N})\over {\nu^2 - (a k + m)^2}} \, , \\
\vspace{0.2cm}
{\cal C}_s &= &  \sin (\nu \tau )\cos ( a k \tau  + m \tau - k \lambda - m \zeta ) 
- { a k + m \over \nu}\{\cos (\nu \tau ) \sin ( a k \tau + m \tau - k \lambda  - m \zeta ) \\
\vspace{0.2cm}
&& + \sin ( k \lambda + m \zeta )\}\, , \\
{\cal D}_s & = & \cos (\nu \tau )\cos (  a k \tau +  m \tau - k \lambda - m \zeta ) 
+ {a k + m \over \nu}\sin (\nu \tau ) \sin ( a k \tau + m \tau - k \lambda - m \zeta )\\
\vspace{0.2cm}
&&  - \cos ( k \lambda + m \zeta) \, , \\
\vspace{0.2cm}
\lambda & = & \phi -  a n \tau \, , \qquad \zeta \; = \; \delta -  n\tau \, ,\\
\tau &= &w_{rot}\bigtriangleup t \, , \qquad n \; = \; 0, 1, 2, 3,......., M - 1.
\end{array} \right\}
\label{eq:sloc}
\end{equation}

\noindent The FT of the complete response would now be given via
\begin{eqnarray}
\label{eq:rfs}
\left[\tilde{R}(f)\right]_s & = & e^{-i 2 \beta_o}\left[\tilde{h}( f
+ 2 f_{rot})/2\right]_s\left[ h_{o_+}( F_{1_+} + i F_{2_+} )
+ h_{o_\times} ( F_{2_\times} - i F_{1_\times} )\right] +\nonumber \\
&& e^{i2\beta_o}\left[\tilde{h}( f - 2 f_{rot})/2\right]_s\left[ h_{o_+}
( F_{1_+} - i F_{2_+} )
- h_{o_\times} ( F_{2_\times} + i F_{1_\times} )\right] +\nonumber \\
& & e^{-i\beta_o}\left[\tilde{h}( f + f_{rot})/2\right]_s\left[ h_{o_+}( F_{3_+} + i F_{4_+} )
+ h_{o_\times} ( F_{4_\times} - i F_{3_\times} )\right] +\nonumber \\
&& e^{i\beta_o}\left[\tilde{h}( f - f_{rot})/2\right]_s\left[ h_{o_+}( F_{3_+} - i F_{4_+} )
- h_{o_\times} ( F_{4_\times} + i F_{3_\times} )\right] +\nonumber \\
&&\left[\tilde{h}(f)\right]_s\left[ h_{o_+}F_{5_+} - i h_{o_\times}F_{5_\times}\right]
\end{eqnarray}

\section{N-component signal}
\label{sec:ftn}
The FT in Eqs.~(\ref{eq:hft}) and~(\ref{eq:rft}) are valid for a
pulsar which emits GW signal at single frequency $f_o$. But there are known 
physical mechanisms which generate GW 
signals consisting of many components. An axially symmetric pulsar
undergoing free precession, emits quadrupole GW at
two frequencies, one equal to the sum of the spin frequency
and the precession frequency, and the other twice of it (Zimmermann and
Szedenits, 1979).
The quadrupole GW from a triaxial ellipsoid rotating
about one of its principal axes consists of one component only (Thorne, 1987). 
In this case the signal has frequency  
twice the spin frequency of the star. In general, if a star is
non-axisymmetric and precesses, the GW signal consists
of more than two components. Recently, new mechanisms e.g.
r-mode instability of spinning neutron stars (Anderson, 1998;
Lindblom et al., 1998; Owen, et al., 1998) and temperature asymmetry in the
interior of the neutron star with miss-aligned spin axis (Bildsten, 1988)
have been discussed in the literature.\\

\par In view of the above discussion CGW signal may 
consist of frequencies which are multiple of some basic frequencies. An
analysis of the GW data of N-component of signal has been made recently by
Jaranowski and Kr\'{o}lak (2000). We do not continue this analysis as the requisite formalism 
is analogous to what we have presented in paper-I and the earlier sections of this paper 

\section{Discussion and conclusions}
\label{sec:concl4}
The analysis and results obtained in paper-I regarding FT of the
response of a Laser Interferometer have been generalised in this
paper. In this context following points must be noted.
\renewcommand{\theenumii}{\roman{enumii}}
\begin{enumerate}
\item For 120 days observation time data the resolution provided by
FFT, which is equal to $1/T_o$, is sufficient to represent the structure of side bands.

\item In every case discussed, it turned out that the maximum power lies
in the frequency $f_o + 2f_{rot}$. However, this is not established conclusively
if this result is generic. 

\item Throughout our analysis in paper-I and this paper we have employed following
conditions.
\begin{enumerate}
\item The phase of the wave is zero at $t = 0$.

\item The observation time of the data set is from $t = 0$ to $t = T_o$.
\end{enumerate}

\item By trivial modifications we could get the general results with non-zero
initial time and phase.
\end{enumerate}

\par It is pointed out that in choosing the parameters for the source and the detectors 
we have been constrained by the computer memory and efficiency. Accordingly the choice of 
the parameters has been a matter of convenience. The application of the 
results obtained in the present and the earlier paper to the problem of all
sky search for continuous gravitational wave sources will be made in our 
forthcoming article (Srivastava and Sahay, 2002 b).

\section*{Acknowledgments}
The authors are thankful to Prof. S. Dhurandhar, IUCAA, Pune for stimulating 
discussions. The authors are extremely thankful to the anonymous referee 
for his hardwork in pinpointing the errors and making the detailed suggestions 
which resulted in the improvement of the paper. The authors are also thankful to 
IUCAA for providing hospitality
where major part of the work was carried out. This work is supported through
research scheme vide grant number SP/S2/0-15/93 by DST, New Delhi.

\newpage

\end{document}